\documentstyle[aps]{revtex}

\newcommand{\ket}[1]{\mbox{$|{#1}\rangle$}}

\begin{document}
\draft
\title{A Practical Method of Constructing Quantum Combinational Logic Circuits}
\author{Jae-Seung Lee, Yongwook Chung, Jaehyun Kim, and Soonchil Lee}
\address{Department of Physics, Korea Advanced Institute of Science and Technology, Taejon 305-701, Korea}
\date{\today}
\maketitle

\begin{abstract}
We describe a practical method of constructing quantum combinational logic circuits with basic quantum logic gates such as NOT and general $n$-bit Toffoli gates. This method is useful to find the quantum circuits for evaluating logic functions in the form most appropriate for implementation on a given quantum computer. The rules to get the most efficient circuit are utilized best with the aid of a Karnaugh map. It is explained which rules of using a Karnaugh map are changed due to the difference between the quantum and classical logic circuits.
\end{abstract}
\pacs{03.67.Lx}

\section{Introduction}
To perform a quantum algorithm, required unitary operations should be expressed as a sequence of the basic operations which can be implemented by a quantum computer. It is proved that the combination of one-qubit operations and controlled-NOT (C-NOT) operations can make any unitary operations \cite{Barenco95}, and the implementation of these basic operations is the least requirement for a quantum system to be a quantum computer. Several methods have been suggested to get an operation sequence, or a quantum circuit, consisting of the basic quantum gates for a given unitary opertaion \cite{Barenco95,Somaroo98,Tucci,Kim}. These methods are general, but they do not necessarily generate the most efficient operation sequence from the point of view of experimental implementation.

The evaluation of a binary or logic function which is the very basic task in classical computation is also frequently required in the quantum algorithms such as search and Deutsch algorithms \cite{Grover97,Deutsch92}. One of the results of studies on quantum computation is that a quantum computer can calculate any function which can be calculated by a classical computer \cite{Bennett89}. Therefore, it is expected that a function can be evaluated in a quantum computer through a {\it quantum combinational logic circuit} corresponding to a classical combinational logic circuit. A quantum combinational logic circuit can be constructed with basic quantum logic gates such as quantum NOT, controlled-NOT (C-NOT) and Toffoli gates as a classical combinational logic circuit is constructed with NOT, AND and OR gates. An efficient method of making a well designed quantum circuit for a logic function will be very useful because the successful implementation of a quantum algorithm depends heavily on the number of basic gates composing the quantum circuit due to the finite coherence time. In fact, this is the most important reason why it is hard to increase the number of qubits in the implementation of quantum computation. In this work, we present a practical method which can make the most efficient operation sequence for the implementation of logic functions in quantum computation, that is, the $f$-controlled NOT and $f$-controlled phase shift gates.

\section{Binary Functions and Quantum Gates}
A classical computer performs combinational and sequential logic operations an assembly of which accomplishes a task. A combinational logic operation is basically the operation of a binary function with an $n$-bit input and a one-bit output:
\begin{equation}
f: \{0,1\}^n \to \{0,1\}.
\label{eqfunc}
\end{equation}
By assembling these operations each of which generates an output of one bit, a computer can evaluate any function with an output of multiple bits. A function can be performed by a combinational logic circuit which consists of NOT, AND, and OR gates. One way of expressing a function is to logically add all minterms defined as
\begin{equation}
f^{(a)} (x) = \left\{ \begin{array}{ll}
1 & x=x^{(a)} \\
0 & \text{otherwise}
\end{array}\right.,
\end{equation}
where $x^{(a)}$ is a constant satisfying $f(x^{(a)})=1$ \cite{Preskill}. Minterms are expressed by logical multiplication of all $n$ bits of an input with NOT operations first applied to the bits of $x^{(a)}$ having 0 as their values. In classical computation, this logical expression can be simplified by using logic identities. Several methods of finding the simplest circuit for a given function are available, including a Karnaugh map \cite{Mano79}.

Because of the reversibility of operations, binary functions cannot be implemented directly in quantum computation. However, they can be realized by unitary transformations as
\begin{equation}
U_{\text{$f$-C-NOT}} \ket{x}\ket{y} = \ket{x}\ket{y \oplus f(x)},
\label{eqfcnot}
\end{equation}
where $f$-C-NOT means $f$-controlled-NOT  \cite{Vedral96}. The input of this operator is stored in the control register
\begin{equation}
\ket{x} = \ket{x_1} \ket{x_2} \ket{x_3} \cdots \ket{x_n} \label{eqconreg}
\end{equation}
composed of $n$ qubits, and the output is given through the function register $\ket{y}$ composed of one qubit \cite{Collins98}.

The $f$-C-NOT gate in Eq.\ (\ref{eqfcnot}) can be easily constructed with the minterm gates which are unitary transformations for $f^{(a)}(x)$'s. Each minterm gate is realized by one generalized $(n+1)$-bit Toffoli gate sandwiched by NOT gates as
\begin{equation}
M^{x^{(a)}}_y \equiv \left( \prod N_{x_i} \right) C^x_y \left( \prod N_{x_i} \right),
\end{equation}
where $N_{x_i}$ and $C^x_y$ represent the NOT gate negating only the state of the $i$-th qubit of the control register and the generalized $(n+1)$-bit Toffoli gate with $n$ control bits $x$ and one target bit $y$, respectively. The product is over all bits of $x^{(a)}$ having 0. The first $\prod N_{x_i}$ is applied for the generalized $(n+1)$-bit Toffoli gate to change the state of the target bit only when $x=x^{(a)}$ and the second $\prod N_{x_i}$ to restore the input to its original state for next use. The $f$-C-NOT gate is obtained by multiplying all its minterm gates as
\begin{equation}
U_{\text{$f$-C-NOT}} = \prod M^x_y,
\label{eqfcnot1}
\end{equation}
where the product is over all $x$'s satisfying $f(x)=1$. Each minterm gate is turned on for only one state of the input, so they all commute with each other and the order in the product can be changed freely. This $f$-C-NOT gate can be simplified as the logic expression is simplified by logic identities. In the following, we describe the rules to find the simplest quantum circuit for an $f$-C-NOT gate.

\section{Rules of Minimization}
In classical computation, a simplified logic circuit can be found by using logic identities. It might be thought that a simplified quantum logic circuit can be also obtained in the same way, that is, by multiplying all quantum gates corresponding to the terms in the simplified logic expression of a function. However, this simple substitution procedure works only for the logic expression composed of minterms because the multiplication of quantum gates is not exactly equivalent to an OR operation in Boolean algebra. For example, the quantum circuit corresponding to a logic expression $x_1 + x_2$ obtained by simple substitution procedure is $C^{x_1}_y C^{x_2}_y$. Since $x_1 + x_2$ gives 0 only when $x_1 = 0$ and $x_2 = 0$, the expression composed of minterms is $x_1 x_2 + \bar{x_1} x_2 + x_1 \bar{x_2}$ where bar means NOT. The corresponding quantum circuit $(C^{x_1 x_2}_y) (N_{x_1} C^{x_1 x_2}_y N_{x_1}) (N_{x_2} C^{x_1 x_2}_y N_{x_2})$ gives correct answers in the function resister $\ket{y}$ while $C^{x_1}_y C^{x_2}_y$ does not because it gives 0 when $x_1 = 1$ and $x_2 = 1$. The reason is that the function resister is flipped twice for this input. Therefore, the rules of minimizing quantum logic circuits are different from those of minimizing classical logic circuits. The number of basic quantum logic gates in a circuit can be reduced using the following rules.

First, if the logic expressions of two $(m+1)$-bit quantum logic gates for $m \le n$ are same except the state of one bit and these two gates can be placed in succession, they can be replaced by one $m$-bit gate (Rule I). For example, two $(n+1)$-bit gates corresponding to minterms $x_1 \cdots x_{i-1} x_i x_{i+1} \cdots x_{n}$ and $x_1 \cdots x_{i-1} \bar{x_i} x_{i+1} \cdots x_{n}$ can be replaced by a new $n$-bit gate corresponding to a logic expression $x_1 \cdots x_{i-1} x_{i+1} \cdots x_{n}$ (Fig.\ \ref{figsimple}(a)). This rule reflects the logic identity $x_i + \bar{x_i} = 1$.

Second, two $(m+1)$-bit quantum logic gates can be replaced by one $m$-bit gate and two C-NOT gates if they can be placed in succession and their logic expressions are same except the states of two bits (Rule II). For example, two $(n+1)$-bit gates corresponding to minterms $x_1 \cdots x_i \cdots \bar{x_j} \cdots x_{n}$ and $x_1 \cdots \bar{x_i} \cdots x_j \cdots x_{n}$ can be replaced by one $n$-bit quantum logic gate corresponding to a logic expression $x_1 \cdots x_{i-1} x_{i+1} \cdots x_j \cdots x_{n}$ sandwiched by two C-NOT gates with a control bit $x_i$ and a target bit $x_j$ (Fig.\ \ref{figsimple}(b)). This rule reflects the logic identities $x_i \oplus x_j = x_i \bar{x_j} + \bar{x_i} x_j$ and $x_i \oplus \bar{x_j} = \bar{x_i} \oplus x_j = x_i x_j + \bar{x_i} \bar{x_j}$.

The number of basic gates can be reduced further by using the fact that the square of a minterm gate equals to the identity, i.e. $(M^x_y)^2=I$ for any $x$. Therefore, $U_{\text{$f$-C-NOT}}$ is unchanged by the multipication of even powers of any minterm gates (Rule III) and reduced further to a simpler form by the Rules I and II. In classical computing, we can add any minterms of $x$ satisfying $f(x)=1$ to the logic expression any number of times and simplify the expression using logic identities. Whereas in quantum computing, we can multiply a quantum logic expression by only even powers of any minterm gates, regardless of whether the minterms are for $x$'s satisfying $f(x)=1$. In fact, the minterms of $x$ satisfying $f(x)=0$ are not less useful than those satisfying $f(x)=1$. As discussed in the next section, a Karnaugh map (K-map) helps to choose useful pairs of minterm gates to be multiplied.

As an example, we will construct an $f$-C-NOT gate for one balanced function $f_b(x)$ with 3-bit inputs. From the truth table of $f_b(x)$ (Table \ref{truthtable}), minterms of $f_b(x)$ are $f^{(\bar{x_1} x_2 \bar{x_3})}(x)$, $f^{(x_1 x_2 \bar{x_3})}(x)$, $f^{(x_1 \bar{x_2} x_3)}(x)$ and $f^{(\bar{x_1} x_2 x_3)}(x)$. So the quantum circuit of $U_{\text{$f_b$-C-NOT}}$ is obtained by multiplying the minterm gates corresponding to these minterms as
\begin{equation}
U_{\text{$f_b$-C-NOT}} = M^{\bar{x_1} x_2 \bar{x_3}}_y M^{x_1 x_2 \bar{x_3}}_y M^{x_1 \bar{x_2} x_3}_y M^{\bar{x_1} x_2 x_3}_y.
\label{eqfbcnot1}
\end{equation}

By Rule I, two 4-bit minterm gates $M^{\bar{x_1} x_2 \bar{x_3}}_y$ and $M^{x_1 x_2 \bar{x_3}}_y$ can be replaced by a 3-bit quantum logic gate
\begin{equation}
L^{x_2 \bar{x_3}}_y = N_{x_3} C^{x_2 x_3}_y N_{x_3}
\end{equation}
which negates the state of the qubit $\ket{y}$ only when $\ket{x_2}=\ket{1}$ and $\ket{x_3}=\ket{0}$ irrespective of the state of qubit $\ket{x_1}$. A minterm gate is the special case of the quantum logic gate $L$ when there are no ``don't care'' qubits, and the generalized Toffoli gate including a C-NOT gate is also the special case of $L$ having no NOT gates. Applying Rule II to $M^{x_1 \bar{x_2} x_3}_y$ and $M^{\bar{x_1} x_2 x_3}_y$, we get
\begin{equation}
M^{x_1 \bar{x_2} x_3}_y M^{\bar{x_1} x_2 x_3}_y = C^{x_1}_{x_2} L^{x_2 x_3}_y C^{x_1}_{x_2},
\end{equation}
and therefore,
\begin{equation}
U_{\text{$f_b$-C-NOT}} = \left( L^{x_2 \bar{x_3}}_y \right) \left(C^{x_1}_{x_2} L^{x_2 x_3}_y C^{x_1}_{x_2} \right).
\label{eqfbcnot2}
\end{equation}
The order of terms grouped by  parentheses can be exchanged because they commute with each other.

If $U_{\text{$f_b$-C-NOT}}$ is multiplied by a pair of minterm gates $(M^{x_1 x_2 x_3}_y)^2$ which are absent in Eq.\ (\ref{eqfbcnot1}) and the gates are shuffled, the logic circuit is simplified by using Rule I as
\begin{eqnarray}
U_{\text{$f_b$-C-NOT}} &=& \left( M^{\bar{x_1} x_2 \bar{x_3}}_y M^{\bar{x_1} x_2 x_3}_y M^{x_1 x_2 \bar{x_3}}_y M^{x_1 x_2 x_3}_y \right) \nonumber \\
& & \times \left( M^{x_1 \bar{x_2} x_3}_y M^{x_1 x_2 x_3}_y \right) \nonumber \\
&=& \left( L^{\bar{x_1} x_2}_y L^{x_1 x_2}_y \right) \left( L^{x_1 x_3}_y \right) \nonumber \\
&=& L^{x_2}_y L^{x_1 x_3}_y
\label{eqfbcnot3}
\end{eqnarray}
where the order of gates can be exchanged as in Eq.\ (\ref{eqfbcnot2}). It is obvious that the circuits of Eq.\ (\ref{eqfbcnot2}) and Eq.\ (\ref{eqfbcnot3}) are simpler and more efficient for implementation than that of Eq\ (\ref{eqfbcnot1}).

One of the logic expressions of $f_b(x)$ simplified by logic identities is $\bar{x_1} x_2 + x_2 \bar{x_3} + x_1 \bar{x_2} x_3$. The quantum circuit obtained by simply substituting the corresponding quantum logic gates for the terms in this simplified logic expression is
\begin{equation}
U' = L^{\bar{x_1} x_2}_y L^{x_2 \bar{x_3}}_y M^{x_1 \bar{x_2} x_3}.
\end{equation}
This is the wrong expression because after expanding $L$'s in terms of minterm gates and using $(M^x_y)^2=1$, we get
\begin{eqnarray}
U' &=& M^{\bar{x_1} x_2 x_3}_y M^{\bar{x_1} x_2 \bar{x_3}}_y M^{x_1 x_2 \bar{x_3}}_y M^{\bar{x_1} x_2 \bar{x_3}}_y M^{x_1 \bar{x_2} x_3} \nonumber \\
&=& M^{\bar{x_1} x_2 x_3}_y M^{x_1 x_2 \bar{x_3}}_y M^{x_1 \bar{x_2} x_3}_y,
\end{eqnarray}
which is obviously different from the correct expression in Eq.\ (\ref{eqfbcnot1}).

\section{Karnaugh maps}
Now the problem is how we can find useful pairs of minterm gates to multiply and group them to get the form best for implementation. We found that a Karnaugh map, which has been used as a method of constructing the  simplest classical circuit for a combinational logic \cite{Mano79,Horowitz}, is also useful for this purpose. In this section, it is explained how a K-map can be used to find the most efficient quantum logic circuit for an $f$-C-NOT gate.

A K-map is a truth table where the variables are represented along two axes. They are arranged in such a way that the state of only one input bit changes in going from one square to an adjacent square. Fig.\ \ref{fig:kmap} shows the K-map for a function with 3-bit input as an example. If two adjacent squares have 1's, two minterms can be grouped and reduced to one logic term having one less input bits (Rule I). If two squares adjacent to a common square have 1's, two minterms are reduced to a logic term including a XOR or a exclusive-NOR (Rule II). The more squares can be grouped, the simpler the logic term becomes. By logically adding all the simplified terms, we get a simplified logic expression for a given function.

Because of the difference between the classical and quantum logic circuits discussed above, however, the rules of grouping squares are somewhat different when groups share common squares. In the classical case, groups include only squares of 1's and each square of 1 can be included repeatedly in any number of groups. But in the quantum case, squares of 0's as well as squares of 1's can be included in groups. By Rule III, it is well understood that the latter must be included in the odd number of groups while the former in the even number of groups. To illustrate the method, we will get the quantum logic expression of the $f$-C-NOT gate of $f_b(x)$ using a K-map as an example.

If we make two non-overlapping groups of 1's as in Fig.\ \ref{figfbkmap}(a), the logic expressions of them are $x_2 \bar{x_3}$ and $(x_1 \oplus x_2) x_3$. The corresponding quantum logic gates are $L^{x_2 \bar{x_3}}_y$ and $C^{x_1}_{x_2} L^{x_2 x_3}_y C^{x_1}_{x_2}$, respectively. By multiplying these two quantum gates, the expression in Eq.\ (\ref{eqfbcnot2}) is obtained. We can group squares so that the square corresponding to the minterm $x_1 x_2 x_3$ for which $f_b(x)=0$ is included in two groups as in Fig.\ \ref{figfbkmap}(b). The logic expressions of the two groups are $x_2$ and $x_1 x_3$ and the corresponding quantum logic gates are $L^{x_2}_y$ and $L^{x_1 x_3}_y$, respectively. Therefore, the expression in Eq.\ ({\ref{eqfbcnot3}) is obtained by multiplying these two quantum gates. This example illustrates the clever choice of the pair of minterm gates to be multiplied to the quantum logic circuit for minimization.

\section{The \lowercase{$f$}-Controlled-Phase-Shift gate}
Our method of minimizing quantum logic circuits is applicable not only to an $f$-C-NOT gate but also to an $f$-controlled-phase-shift ($f$-C-PS) gate
\begin{equation}
U_{\text{$f$-C-PS}} \ket{x} = (-1)^{f(x)} \ket{x},
\label{eqfphase}
\end{equation}
required in some algorithms such as Grover's search algorithm \cite{Grover97}. It is well known that the control register of an $f$-C-NOT gate in Eq.\ (\ref{eqfcnot}) is transformed as the variable in Eq.\ (\ref{eqfphase}) if the function register $\ket{y}$ is initially prepared in the superposition state $\frac{1}{\sqrt{2}} ( \ket{0} - \ket{1} )$, that is,
\begin{equation}
U_{\text{$f$-C-NOT}} \ket{x} \frac{1}{\sqrt{2}} ( \ket{0} - \ket{1} ) = (-1)^{f(x)} \ket{x} \frac{1}{\sqrt{2}} ( \ket{0} - \ket{1} ).
\end{equation}
Here, an $(n+1)$-bit $f$-C-NOT gate operates like an $n$-bit $f$-C-PS gate since the state of the function resister remains same after the gate operation. Therefore, it is expected that the quantum circuit of an $n$-bit $f$-C-PS gate can be constructed in the similar way as that of an $(n+1)$-bit $f$-C-NOT gate is constructed.

Let $R^x$ be the quantum phase shift gate which changes the sign of the control resister only when the states of qubits make the value of the logic expression of $x$ equal to 1. The gate $R^{x}$ can be constructed as $H_{x_i} L^{x_1 \cdots x_{i-1} x_{i+1} \cdots x_n}_{x_i} H_{x_i}$ for any $x_i$, where $H_{x_i}$ is a Hadamard gate operating on $x_i$ \cite{Gershenfeld97,Jones98-3}. The gate $R^x$ has the same property as the logic gate $L^x_y$ in the sense that they are active only when the control register satisfies the condition that $x=1$ and squares of them are equal to the identity. Therefore, the rules to get the simplified circuit for an $f$-C-NOT gate are also applicable for an $f$-C-PS gate.

Consequently, the quantum logic expression of an $n$-bit $f$-C-PS gate can be obtained from that of an $(n+1)$-bit $f$-C-NOT gate by directly replacing all $L$'s which take the function register as their target bits by corresponding R's. For example, the quantum circuit of the $f$-C-PS gate corresponding to $f_b(x)$ are given as
\begin{eqnarray}
\lefteqn{U_{\text{$f_b$-C-PS}}} \nonumber \\
&=& R^{\bar{x_1} x_2 \bar{x_3}} R^{\bar{x_1} x_2 x_3} R^{x_1 \bar{x_2} x_3} R^{x_1 x_2 \bar{x_3}}
\label{eqfbphase1} \\
& = & \left( R^{x_2 \bar{x_3}} \right) \left( C^{x_1}_{x_2} R^{x_2 x_3} C^{x_1}_{x_2} \right) \label{eqfbphase2} \\
& = & R^{x_2} R^{x_1 x_3} \label{eqfbphase3}.
\end{eqnarray}
The expressions of Eq.\ (\ref{eqfbphase1}), (\ref{eqfbphase2}) and (\ref{eqfbphase3}) were obtained from Eq.\ (\ref{eqfbcnot1}), (\ref{eqfbcnot2}) and (\ref{eqfbcnot3}), respectively.

\section{Conclusion}
The most useful aspect of this method is that it helps to find the quantum circuit most appropriate for a given quantum computer. For example, when a quantum system consists of qubits some of which have no or weak interactions between them, swap gates, each of which consists of three C-NOT gates \cite{Preskill}, are used to get the same effect as the gates requiring those interactions \cite{Lloyd93}. However, the usage of swap gates makes the operation sequence composed of basic logic gates much longer. In this case, our method can give the quantum circuit which do not require those interactions. Suppose $U_{\text{$f_b$-C-PS}}$ should be implemented on the quantum computer in which there is no interaction between qubits corresponding to the control bits $x_1$ and $x_3$. Then the logic expression of Eq.\ (\ref{eqfbphase2}) is most suitable for this purpose because it does not contain gates requiring the interaction between qubits $\ket{x_1}$ and $\ket{x_3}$. The expression of Eq.\ (\ref{eqfbphase3}) looks most simple, but it requires two swap gates to implement the gate $R^{x_1 x_3}$. The minimized quantum circuit does not necessarily mean the most efficient implementation.

We proposed a method useful for constructing the most efficient quantum combinational logic circuit for evaluation of a logic function on a given quantum computer. The rules for simplification can be utilized best with the aid of a K-map. The K-map with new rules for the quantum logic circuit is the powerful visual tool especially for the circuits of binary functions with 3- and 4-bit inputs.

\begin{figure}[p]
\caption{(a) A quantum circuit composed of two minterm gates (left) and the circuit simplified by Rule I (right). After the operation, the function register $\protect\ket{y}$ carries $\protect\ket{y \oplus x_1 \cdots x_{i-1} x_{i+1} \cdots x_n}$. (b) A quantum circuit composed of two minterm gates (left) and the circuit simplified by Rule II (right). After the operation, the function register $\protect\ket{y}$ carries $\protect\ket{y \oplus [ (x_i \oplus x_j ) ( x_1 \cdots x_{i-1} x_{i+1} \cdots x_{j-1} x_{j+1} \cdots x_n) ]}$.}
\label{figsimple}
\end{figure}

\begin{figure}
\caption{The Karnaugh map of a 3-bit input binary function.}
\label{fig:kmap}
\end{figure}

\begin{figure}
\caption{K-map's used for construction of the quantum circuit of $U_{\text{$f$-C-NOT}}$ (left) and the corresponding quantum circuits (right). (a) Two non-overlapping groups of 1's are chosen. (b) One large and one small groups share a square of 0. The squares of 0 must belong to the even numbers of groups.}
\label{figfbkmap}
\end{figure}

\begin{table}
\caption{The truth table of a 3-bit binary function $f_b(x)$.}
\label{truthtable}
\begin{tabular}{ccc|c}
$x_1$ & $x_2$ & $x_3$ & $f_b (x)$ \\ \tableline
0 & 0 & 0 & 0 \\
0 & 0 & 1 & 0 \\
0 & 1 & 0 & 1 \\
0 & 1 & 1 & 1 \\
1 & 0 & 0 & 0 \\
1 & 0 & 1 & 1 \\
1 & 1 & 0 & 1 \\
1 & 1 & 1 & 0 \\
\end{tabular}
\end{table}
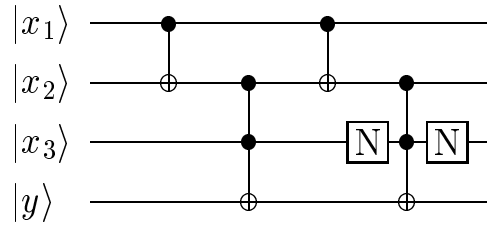
\end{document}